# Design guidelines for a highly efficient high-purity Germanium (HPGe)-based double-heterojunction solar cell

Jaker Hossain[a,*], Md. Mahabub Alam Moon[b], Bipanko Kumar Mondal[a] and Mohammad Abdul Halim[c]

[a]Solar Energy Laboratory, Department of Electrical and Electronic Engineering, University of Rajshahi, Rajshahi 6205, Bangladesh.

[b]Department of Electrical and Electronic Engineering, Begum Rokeya University, Rangpur, Rangpur 5400, Bangladesh.

[c]Department of Materials Science and Engineering, Faculty of Engineering, University of Rajshahi, Rajshahi 6205, Bangladesh.

**Abstract**

In spite of having higher carrier mobilities and absorption coefficients of germanium (Ge) than those of silicon (Si), there has been less focus on Ge-based solar cells due to the low bandgap and high-cost of Ge wafer as well as requirement of its high-purity level. Currently, availability of high-purity Ge (HPGe), the low-cost wafer slicing method and proper design guidelines make it possible to design HPGe-based solar cells. Accordingly, in this article, we have designed and simulated a novel $n$-CdS/$p$-HPGe/$p^+$-BaSi$_2$ based $npp^+$ double-heterojunction solar cell, where HPGe, cadmium sulfide (CdS) and orthorhombic barium disilicide (β-BaSi$_2$) have been used as the absorber, window and back-surface field (BSF) layers, respectively. Using the solar cell capacitance simulator (SCAPS-1D), the effects of different physical parameters such as the thickness, doping and defect densities, band offsets and temperature on the photovoltaic (PV) parameters of the designed solar cells have been investigated systematically. This article renders the optimized PV parameters to improve the device performance with the highest power conversion efficiency (PCE) of ~45.65% with a high open-circuit voltage of 1.16 V owing to the high built-in voltage of 1.7 V for the $n$-CdS/$p$-HPGe/$p^+$-BaSi$_2$ solar cells. This efficiency is almost consistent with the detailed-balance limit for double heterojunction solar cell.

*Keywords:* High-purity Germanium, BaSi$_2$ back-surface field, Built-in potential, Bulk defect, Heterojunction solar cell, SCAPS-1D simulation.

## 1. Introduction

The importance of the study of photovoltaics (PV) as a renewable energy technology has become more and more apparent due to the global energy crisis and green-house effect. Crystalline silicon (c-Si) solar cells have been dominating the PV technology since the last few decades with over 90% share of the global market. However, it requires a very high-temperature of ~1400 °C and high-vacuum fabrication process (Andreani et al., 2019; Battaglia et al., 2016). A recent report shows a power conversion efficiency (PCE) of 26.33% for this solar cell with a practical module size of 180 cm$^2$ using back contact device structure (Yoshikawa et al., 2017).

PV researchers are trying to make alternative cheap and efficient solar cells to meet the increasing energy demand. Over the last few years, a number of new solar cells structures such as Si/Organic (Avasthi et al., 2011; Zielke et al., 2014; Funda et al., 2016; Kasahara et al., 2018), Si/Chalcogenide (Mondal et al., 2019), Perovskites (Kayesh et al., 2019; Zhu et al., 2019; Hou et al., 2020), Silicide-based (Moon et al., 2020a, 2020b; Biplab et al., 2019) solar cells etc. have already been designed and demonstrated. More recently, Moon et al. (2020b) have reported a highly promising iron disilicide (FeSi$_2$)-based solar cell with a high PCE of ~39%, where FeSi$_2$ having a low bandgap of 0.87 eV has been used as the absorber layer. Despite such a low bandgap of the absorber, the open-circuit voltage of ~0.96 V has been achieved by the use of heavily doped cadmium sulfide (CdS) and barium disilicide (BaSi$_2$) as the window and back-surface field (BSF) layers, respectively.

Herein, we propose a very high-efficiency novel solar cell structure based on high-purity germanium (HPGe). Germanium (Ge) is a group IV indirect semiconductor like Si, but with a smaller bandgap of 0.66 eV and higher carrier mobilities at low temperature, which make it more attractive than Si for optoelectronic applications (Dalven, 1966; Debye and Conwell, 1954). It is worth noting here that relatively low-purity Ge (LPGe) crystals have so far been used in wave guide, infrared (IR) window fiber optics, IR night vision devices, space solar cells and as polymerization catalyst (Wada and Kimerling, 2015), whereas the HPGe crystals have the applications in the fabrication of nuclear and gamma ray detectors (Hansen and Haller, 1981; Haller et al., 1979, 1972; Eberth and Simpson, 2008; Yang et al., 2012).  However, a very few works have been performed on Ge as a solar cell (Sorianello et al., 2013). Burnett (2002) has reported the simulation study of Si/LPGe tandem solar cell with a predicted efficiency of 19-28%. In addition, LPGe is used as the bottom layer of the multijunction solar cells, where gallium-indium-arsenide (GaInAs) and gallium-indium-phosphide (GaInP) layers are used at the top of this layer (Sun et al., 2010).

Generally, bulk defects in the absorber layer of a solar cell degrade the performance of the device to a great extent. Therefore, a control of the grown-in defects is prerequisite for the finest performance. Multicomponent-based semiconductor absorbers have possibilities to form different types of defects in a wide range such as vacancies, antisites, secondary phases etc., which limit the PCE of the solar device substantially. As a result, the growth of such kind of high-purity multicomponent layer remains very far to the PV researchers. On the other hand, detector grade HPGe with very low elctrically active impurity concentration is grown by Czochralski (CZ) method, an well established technology like that of silicon, in hydrogen atmosphere in which dislocation density ranges in $2\times10^3$ to $7\times10^3$ cm$^{-2}$ (Wang et al., 2014). Blondeel and Clauws (1999) have described that the copper-related defects which are introduced into the p-type HPGe by diffusion into the samples during heat treatment are in the range from $10^6$ to $10^8$ cm$^{-3}$.

Ge-based new concepts of materials and devices are becoming popular due the development of germanium oxynitride as a mask material and the tuning of its bandgap by band-structure engineering (Chui et al., 2006; Gupta et al., 2013; Pillarisetty, 2011; Collaert et al., 2015). However, the cost of these devices was high due to the expense of high-quality Ge wafers. Thanks to recent technology, a low-cost and flexible single-crystalline-like Ge template on flexible metal substrates has already been developed by the ion beam-assisted deposition (IBAD) technique (Selvamanickam et al., 2009) for use in high-performance Si flexible electronics (Gao et al., 2016, 2014) and gallium arsenide (GaAs)-based solar cells (Dutta et al., 2014; Vanhellemont et al., 2007). In addition, Bamberg and Rakwal (2008) have introduced a new wafer slicing technique, namely the wire electrical discharge machining (WEDM) used to reduce the cost as it minimizes wastage in Ge cutting and produces even thinner wafers with less cracking. More recently, Li et al. (2018)

have reported that HPGe can be grown with a dislocation density of ∼5×10$^6$ cm$^{-2}$ by liquid phase epitaxy (LPE), which offers a high-growth rate that is beneficial for low-cost mass production.

In addition to the HPGe thin film fabrication technique, concerning the growth of a thin layer of BaSi$_2$, techniques including vacuum evaporation (Hara et al., 2016), molecular beam epitaxy (MBE) (Deng et al., 2018; Du et al., 2015), solid phase epitaxy (SPE) (Fomin et al., 2017) etc. have been reported by this time. Besides, Yoneyama et al. (2013) have employed the radio frequency (RF) magnetron sputtering method to grow polycrystalline BaSi$_2$ films on glass substrate in a low-cost lagre-scale high-speed process. Regarding the synthesis of highly transparent CdS thin film as the window layer for solar cell applications, we have recently reported the low-cost sol-gel spin coatong method incorporating Triton-X 100 (TX-100) surfactant (Rahman et al., 2020a, 2020b). Among the other deposition techniques, chemical bath deposition (CBD), chemical vapor deposition (CVD), electrodeposition, MBE, spray pyrolysis, pulsed DC magnetron sputtering, electron beam evaporation etc. can be mentioned (Rahman et al., 2020a).

To the best of our knowledge, for the first time, we concentrate on the application of HPGe as a novel photonic material for thin film absorber layer in solar cell structure. In this work, we design and simulate HPGe-based double-heterojunction solar cells (HJSCs) of the structure: *n*-CdS/*p*-HPGe/*p*$^+$-BaSi$_2$ that incorporate heavily doped BaSi$_2$ as the BSF and CdS as the window materials, respectively. The effect of different physical parameters on the photovoltaic performances such as the short-circuit current density ($J_{SC}$), open-circuit voltage, ($V_{OC}$), fill factor (FF), PCE ($\eta$) and quantum efficiency (QE) have been investigated for the designed solar cells. This study reveals that this newly designed HPGe-based *npp*$^+$ heterojunction thin film solar cell (TFSC) is highly promising for large-scale applications in space and multi-kilowatt systems of the PV industries.

## 2. Device structure and simulation

Fig. 1 (a) shows the schematic structure of the *n*-CdS/*p*-HPGe/*p*$^+$-BaSi$_2$ HJSC. In the solar cell layout, HPGe, BaSi$_2$ and CdS are used as the active *p*-type absorber layer, *p*$^+$-type BSF layer and *n*-type window layer, respectively. Simulation of the modeled solar cell was performed under global air mass (AM) 1.5 spectrum at a one sun illumination (100 mW cm$^{-2}$) at a temperature (*T*) of 300 K. Both the single -acceptor and -donor like bulk defects were distributed in the bulk layer using the Gaussian energetic distribution. The values of the surface work function (WF) of the front and rear contacts were collected from the book (Lide, 2005). Materials optoelectronic parameters (Table 1) have been used as the input parameters during simulations.

**Table 1**: Values of input parameters used in the simulation of the $n$-CdS/$p$-HPGe/$p^+$-BaSi$_2$ solar cell at 300 K.

| Parameters | CdS window[a] | HPGe absorber[b] | BaSi$_2$ BSF[a] |
|---|---|---|---|
| Majority carrier | electron | hole | hole |
| Layer thickness, $d$ [µm] | 0.10 | **2.00** | **0.10** |
| Band gap, $E_G$ [eV] | 2.4 | 0.66 | 1.3 |
| Electron affinity, $\chi$ [eV] | 4.2 | 4.00 | 3.3 |
| Dielectric permittivity, $\varepsilon$ [relative] | 10 | 16.20 | 15 |
| Effective conduction band density of states, $N_C$ [cm$^{-3}$] | $2.2 \times 10^{18}$ | $1.05 \times 10^{19}$ | $2.6 \times 10^{19}$ |
| Effective valence band density of states, $N_V$ [cm$^{-3}$] | $1.8 \times 10^{19}$ | $3.92 \times 10^{18}$ | $2 \times 10^{19}$ |
| Mobility (electron), $\mu_n$ [cm$^2$ V$^{-1}$ s$^{-1}$] | 100 | 3900 | 20 |
| Mobility (hole), $\mu_p$ [cm$^2$ V$^{-1}$ s$^{-1}$] | 25 | 1900 | 20 |
| Donor density, $N_D$ [cm$^{-3}$] | $10^{18}$ | 0 | 0 |
| Acceptor density, $N_A$ [cm$^{-3}$] | 0 | $10^{17}$ | **$10^{20}$** |
| Type of defect | Acceptor | Donor | Donor |
| Energetic distribution | Gaussian | Gaussian | Gaussian |
| Defect density, $N(t)$ [cm$^{-3}$] | $10^{14}$ | **$10^6$** | $10^{14}$ |
| Reference energy [eV] | 1.2 | 0.33 | 0.65 |
| Capture cross section of electron for acceptor defect [cm$^2$] | $10^{-15}$ | - | - |
| Capture cross section of hole for acceptor defect [cm$^2$] | $10^{-15}$ | - | - |
| Capture cross section of electron for donor defect [cm$^2$] | - | $10^{-15}$ | $10^{-15}$ |
| Capture cross section of hole for donor defect [cm$^2$] | - | $10^{-17}$ | $10^{-15}$ |
| Reference: a = (Biplab et al., 2019; Moon et al., 2020a, 2020b); b = ([dataset]) | | | |

Note: Values in Bold are variable.

In this work, a one-dimensional solar cell capacitance simulator (SCAPS-1D) of version 3.3.07 (Burgelman et al., 2000; Decock et al., 2012) was used to conduct the simulation and analysis numerically. Although SCAPS simulator enables us to analyze the device having up to seven layers, present structure of the modeled solar cell with only three layers have significance in the photovoltaic performances as will be described in Section 3. Using the abovementioned simulation tool, we computed the band diagram, current density-voltage (*J-V*) characteristics, capacitance-voltage (*C-V*) characteristics, and spectral responses (SR) at light.

## 3. Results and discussion

*3.1. Energy band diagram and built-in-potential of the simulated solar cell*

The modeled $n$-CdS/$p$-HPGe/$p^+$-BaSi$_2$ solar cell is illuminated under a one sun illumination of 100 mW cm$^{-2}$ at global air mass AM 1.5G spectrum at 300 K. Fig. 1 (b) shows the energy band diagram (EBD) of the proposed solar cell under illumination. As seen from the figure, the fermi energy level ($E_F$) enters into the valence band (VB) top edge ($E_V$) of the heavily doped ($10^{20}$ cm$^{-3}$) $p^+$-BaSi$_2$ BSF layer, resulting in a layer of degenerate type semiconductor (Moon et al., 2020b). Since the VB offset is much smaller than the conduction band (CB) offset between the BSF and the absorber layers, BaSi$_2$ and the back (+ve) electrode will easily extract the holes (+ve carriers) from the HPGe absorber. In contrast, the holes travelling to the CdS window layer will be prevented by the large VB offset between the absorber and CdS window layers.

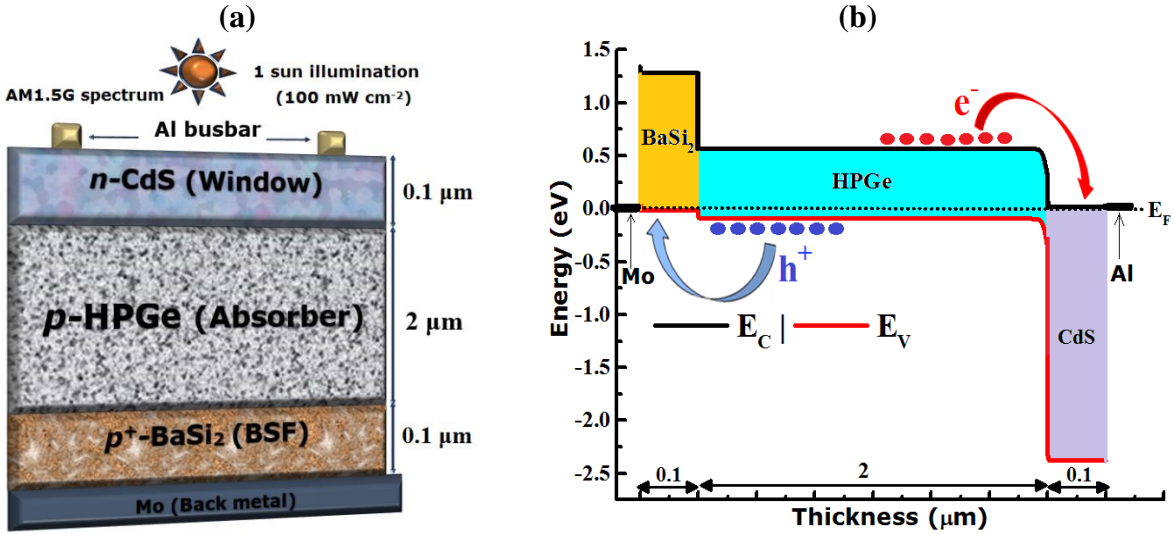

**Fig. 1.** The (a) schematic structure and (b) energy band diagram of the proposed $n$-CdS/$p$-HPGe/$p^+$-BaSi$_2$ double-heterojunction solar cell.

There are two heterojunctions ($p$-$n$ and $p^+$-$p$) formed in the solar cell with the doping densities: $N_A$ in $p$-absorber, $N_D$ in $n$-window and $N_a$ in $p^+$-BSF layers, respectively. The built-in potential ($\psi_{p-n}$) of the hetero $p$-$n$ junction can be found from the following Eq (1) (Wolfe et al., 1989),

$$\psi_{p-n} = \frac{\Delta E_C - \Delta E_V}{2q} + \frac{KT}{q} \ln \frac{N_D N_A}{n_{i_n} n_{i_p}} + \frac{KT}{2q} \ln \frac{N_{v_n} N_{c_p}}{N_{v_p} N_{c_n}} \qquad (1)$$

Where, $N_C$ and $N_V$ are the effective density of states (DOS) at CB and VB edges, $KT$ is the thermal voltage, $q$ is the charge of an electron and $n_{i_n}$ and $n_{i_p}$ are the intrinsic carrier concentration of the corresponding n-type CdS ($1.19 \times 10^{19}$ cm$^{-3}$) (Kose et al., 2010) and p-type Ge ($2.33 \times 10^{13}$ cm$^{-3}$) (Madelung et al., 2002) layers, respectively. For the hetero $p^+$-$p$ junction, the built-in potential ($\psi_{p^+-p}$) can be found as Eq (2) (Wolfe et al., 1989),

$$\psi_{p^+-p} = \frac{\Delta E_C - \Delta E_V}{2q} + \frac{KT}{q} \ln \frac{N_a N_A}{n_{i_p} n_{i_{p+}}} + \frac{KT}{2q} \ln \frac{N_{v_p} N_{c_{p+}}}{N_{v_{p+}} N_{c_p}} \qquad (2)$$

Where, $n_{i_{p+}}$ is the intrinsic carrier concentration of the corresponding $p^+$-type BaSi$_2$ ($10^{16}$ cm$^{-3}$) layer (Du et al., 2014). Consequently, the built-in potential ($\psi$) for the *n*-CdS/*p*-HPGe/$p^+$-BaSi$_2$ heterojunction solar cell can be written as Eq (3),

$$\psi = \psi_{p-n} + \psi_{p^+-p} \qquad (3)$$

The calculated built-in potentials of the *n-p*: CdS-HPGe, *p-p$^+$*: HPGe-BaSi$_2$ and *n-p-p$^+$*: CdS-HPGe-BaSi$_2$ heterojunction solar cells are ~1.05, ~0.65 and ~1.7 V, respectively. This high built-in voltage of 1.7 V comes from the double heterojunction layout designed herein. A very close built-in potential of 1.5 V has been reported by Du et al. (2014) for the n-BaSi$_2$/p-Si single-heterojunction.

*3.2. Effect of the thickness of HPGe absorber layer on the photovoltaic parameters*

The thickness (*d*) of HPGe absorber layer was varied from 0.1-5 µm, keeping all other parameters constant as mentioned in Table 1 at *T* = 300 K. Fig. 2 shows the effect of thickness on the solar cell performance parameters for the said range of thickness variation.

*3.2.1. Effect on $J_{SC}$*

Ge has a very low bandgap energy of 0.66 eV, which implies a very high cutoff wavelength of ~1877 nm in the solar spectrum for absorbing photons. Even this material offers a very high value of absorption coefficient (α) of $6.28 \times 10^3$ cm$^{-1}$ at the longer wavelength of 1.2 µm (Dash and Newman, 1955; Philipp and Taft, 1959), whereas at this wavelength Si hardly absorbs any photon. This contributes to add extra advantages for Ge over Si. As a result, to absorb photons till 1.2 µm spectrum, Ge layer needs the absorption depth (1/α) of only 1.59 µm. Thus, a thickness (*d*) of 2 µm can be considered to be enough to capture all the incident photons from sunlight till the cutoff wavelength.

In this work, we also observe that the $J_{SC}$ increases from 27.88 mA/cm$^2$ at $d = 0.1$ µm to 46.84 mA/cm$^2$ at $d = 2$ µm. The increase of current with increasing absorber layer thickness is reasonable because a thicker absorber will absorb more photons incident on to it and create more electron-hole pairs (EHPs) (Moon et al., 2020a; Biplab et al., 2019). On the contrary, a lower current for a thinner absorber is due to the higher recombination rate of the charge carriers before they reach the *p-n* junction (Moon et al., 2020b) and a lower probability of absorbing photons (Moon et al., 2020a). When the thickness is $d \geq 2$ µm, $J_{SC}$ is almost constant at 49 mA/cm$^2$. So, a thickness of 2 µm is chosen for the absorber layer to move on to investigate further.

*3.2.2. Effect on $V_{OC}$*

When the absorber is thicker, more photons implying more energies are absorbed and the cell produces a higher $V_{OC}$ (Moon et al., 2020b). It has been noticed in the case of FeSi$_2$-based solar cell that the $V_{OC}$ was in a decreasing mode with increasing absorber thickness (Moon et al., 2020b). However, as shown in Fig. 2 in our case of present study, $V_{OC}$ increases with an increase in thickness with a slight increment of only 10 mV from 1.15 V at $d = 0.1$ µm to 1.16 V to $d = 5$ µm. The discrepancy between the previous (Moon et al., 2020b) and present cases can be attributed to the possible least number of defects of 10$^6$ cm$^{-3}$ in the HPGe absorber.

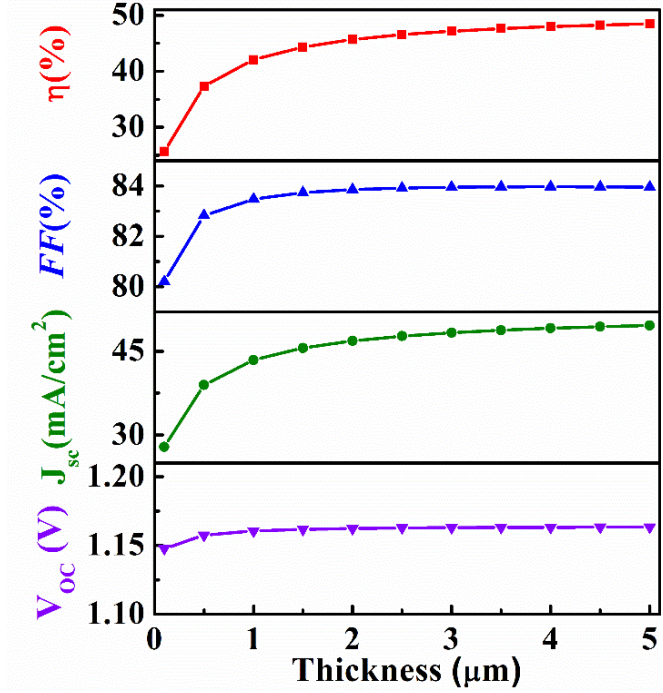

**Fig. 2.** Effect of HPGe layer thickness on the solar cell performance parameters.

However, the calculated value of built-in potential, $\psi$ of the modeled solar cell is ~1.7 V, which is at least twice the bandgap energy of 0.66 eV of HPGe layer. This finding also agrees well with the works reported earlier (Du et al., 2014; Moon et al., 2020b). It has been reported that $\psi$ can be greater than $E_G$ of the absorber when there is degenerate semiconductors as junction partner materials (Gao et al., 2011; Moon et al., 2020b). In the BSF layer the doping density was $10^{20}$ cm$^{-3}$ which produces a higher $\psi$ in the solar cell. This is also reasonable and agrees well with another of our works (Mondal et al., 2019) that more the doping density, higher is the built-in potential. As described in Section 3.1, such a voltage of 1.7 V of HPGe-based solar cell is the sum of the voltages built-in the two junctions. Similar to the case of $\psi > E_G$ for the solar cell structure (Moon et al., 2020b), currently obtained $V_{OC}$ of 1.16 V is within its maximum $\psi$ limit of 1.7 V. It is worth noting here that a very high $V_{OC}$ of 1.16 V from HPGe-based solar cell is feasible if and only if the defects present in HPGe layer is maintained within a level of $10^4$ to $10^7$ cm$^{-3}$ limit, which will be analyzed in a later Section 3.3.

*3.2.3. Effect on FF and PCE*

The parameter FF depends fully on the four paramters *viz.*, the two currents at short-circuit and maximum power point (MPP) condition and the two voltages at open-circuit and MPP condition.

Fig. 2 shows that the FF increases from 80.20% to 83.95% for increasing thickness $d$ from 0.1-5 μm because both the $V_{OC}$ and $J_{SC}$ increase. Since the change in $V_{OC}$ compared to the change in $J_{SC}$ is not significant with respect to the change in $d$, the curve of FF is in consort with the curve of $J_{SC}$.

The PCE is the ratio of the electrical output power from the solar cell to the solar radiation power incident to that cell. Based on the parameters already described above, PCE increases from 25.67% to 48.45% for increasing $d$ from 0.1-5 µm. Such high PCEs are fair due to the high photo-generated current of greater than 46 mA/cm$^2$ at $d > 2$ µm, along with a very high $V_{OC}$ of 1.16 V, produced by the HPGe-based $npp^+$ double heterojunction solar cell.

*3.2.4. Effect on QE*

The QE as a function of light wavelength ($\lambda$) is defined as the ratio of current passing through the circuit *i.e.*, the number of charge carriers taken out from the solar cell to the number of incident photons. When an EHP is produced by each incident photon of a specific wavelength, we could have a 100% external QE (EQE) for that wavelength (Biplab et al., 2019; Moon et al., 2020b).

Fig. 3 shows the effect of varying absorber layer thickness on the QE spectrum. The lower and higher parts of the QE($\lambda$) spectrum depend on the diffusion length of the CdS window and HPGe absorber layer, respectively (Moon et al., 2020b). First, the QE of the CdS/HPGe/BaSi$_2$ solar cell increases with increasing HPGe absorber layer thickness, because a thicker absorber can capture more photons. Then, all the curves at certain wavelengths tend to decrease towards 0% QE depending on its absorption probability in a particular wavelength region of the incident light spectrum. The result matches well with that of the literature (Biplab et al., 2019; Moon et al., 2020b). The QE becomes 0% for the photon energy ($h\nu$) below the band gap ($E_G$) of the HPGe absorber. Other information regarding the QE spectrum has already been discussed in Section 3.2.1.

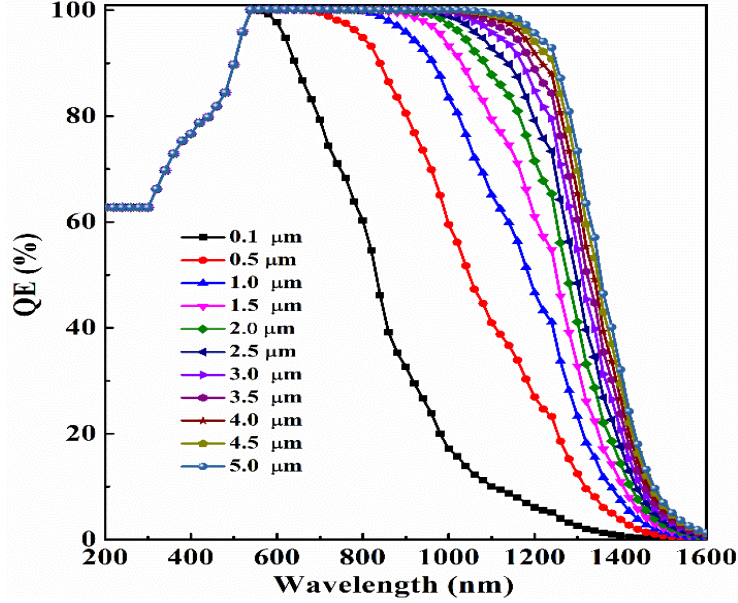

**Fig. 3.** QE($\lambda$) variation with respect to HPGe thickness.

Finally, based on the above discussion and as percieved from Fig. 2 and 3, choosing a thickness of 2 μm of HPGe absorber layer is reasonable.

*3.3. Influence of the bulk defect density of HPGe on solar cell performance*

To investigate the influence of bulk defects of HPGe absorber layer on the output performance parameters of the solar cell, we varied the defect density of the absorber layer from $10^4$-$10^{10}$ cm$^{-3}$ at $T$ = 300 K, keeping all other parameters unchanged as specified (Table 1). The assumed defects were single -acceptor and -donor type in the *n*-CdS window and *p*-HPGe (or *p*$^+$-BaSi$_2$) layers, respectively. However, the interface is assumed undefected in this study. Fig. 4 shows the effect of bulk defect density on the output. It is found that all parameters deteriorate drastically with increasing number of defects when the defect density exceeds $10^7$ cm$^{-3}$. At a higher defect density, the charge carriers recombine timelier with the opposite charge carriers before they reach the junction (Moon et al., 2020a; Biplab et al., 2019). Again, higher the defect density, lower is the rate of creating EHPs.

As shown in Fig. 4, FF decreases in a manner fully based on $V_{OC}$ because $J_{SC}$ remains almost constant for the entire variation of defect density. The $V_{OC}$ drops from 1.16 to 0.91 V while $J_{SC}$ remains unchanged at 46.84 mA/cm$^2$. Thus, the PCE reduces from 45.65% at a defect density of $10^6$ cm$^{-3}$ to 31.54% at a defect density of $10^{10}$ cm$^{-3}$. Unlike today's solar cell absorbers, HPGe is right away availabe in the market with an impurity level of $10^6$-$10^8$ cm$^{-3}$ (Blondeel and Clauws, 1999). This offers great advantages over other solar cell technologies currently in practice. Finally, to maintain a promising result from this newly designed HPGe-based solar cells, it is prerequisite to keep the defect density level of the material within the limit of $10^6$ cm$^{-3}$.

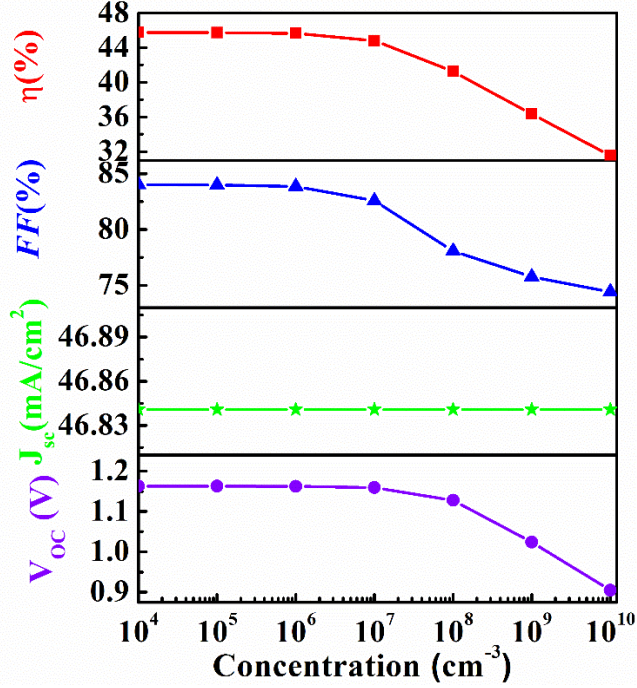

**Fig. 4.** PV parameters variation with bulk defects in HPGe.

*3.4. Impact of the thickness of BSF layer on device performance*

Fig. 5 shows the variation of the PV parameters with respect to the thickness of the BaSi$_2$ BSF layer. It is seen that all parameters excluding the current deteriorate drastically with increasing BSF layer thickness. The current remains approximately the same because the 2 μm thick absorber absorbs almost all the incident photons. Thus, increasing BSF layer thickness does not contribute to photon absorption and thus $J_{SC}$. The dark saturation current density ($J_0$) on the other side increases rapidly with increasing BSF layer thickness, which is confirmed by the dark $J$-$V$ characteristics during the simulations. As $J_{SC}$ remains almost constant and $J_0$ increases, $V_{OC}$ reduces consequently as given by Eq (4) (Moon et al., 2020b),

$$V_{OC} \approx \frac{K_B T}{q} ln \frac{J_{SC}}{J_0} \qquad (4)$$

The decreased $V_{OC}$ thus reduces the FF and PCE to some extent as shown in Fig. 5.

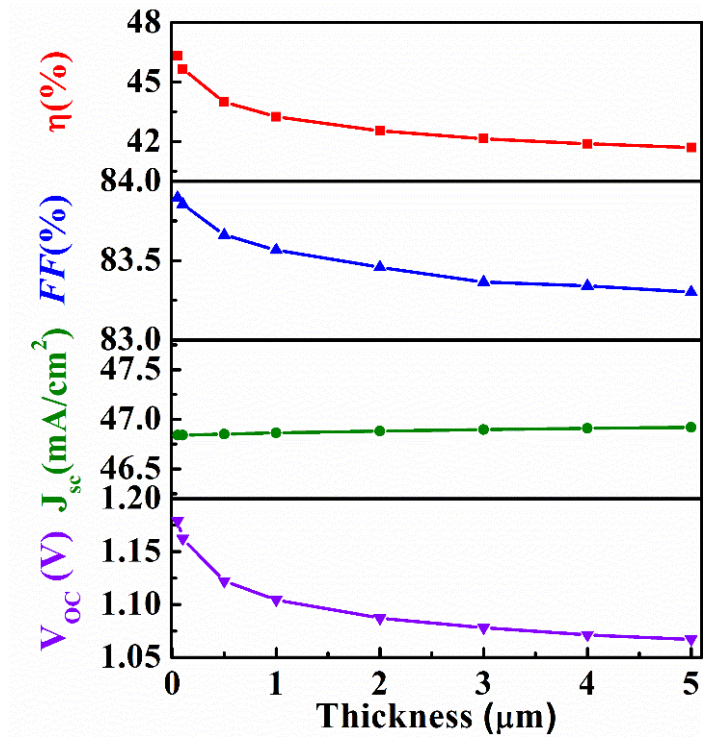

**Fig. 5.** Variation of PV parameters as a function of the thickness of BaSi$_2$ BSF layer.

Fig. 6 shows the variation of QE(λ) as a function of BSF layer thickness. When a BSF layer is introduced in the HJSC, QE increases by a little degree at the high wavelength region of the QE spectrum (Moon et al., 2020b). This is also similar to the result of BaSi$_2$ BSF-aided CIGS-based solar cell (Biplab et al., 2019). However, the QE is found independent of the thickness of the BSF layer, which can be visualized from Fig. 6. Although the presence of a heavily doped BSF layer improves the charge collection probability, the charge collection probability does not depend on its thickness. This is why a thicker BSF layer does not increase the QE.

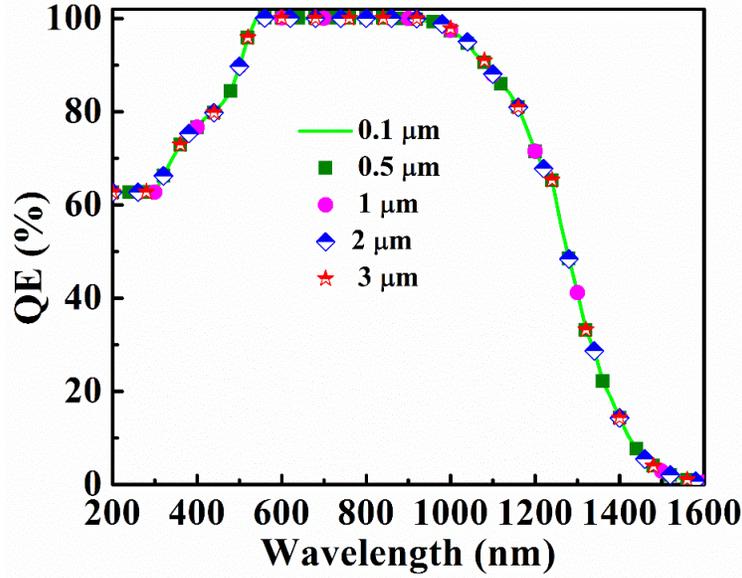

**Fig. 6.** QE variation with the thickness BaSi$_2$ BSF layer.

*3.5 Effect of doping density of the BSF layer*

In this section, we have analyzed the effect of doping level of the BaSi$_2$ BSF layer on the solar cell parameters. BaSi$_2$ has a lifetime of 14 μs and long minority carrier diffusion length of 10 μm (Moon et al., 2020a). Fig. 7 shows the results obtained when the density of the acceptor ions ($N_A$) was varied from $10^{18}$-$10^{22}$ cm$^{-3}$ at $T = 300$ K, taking all other parameters as given in Table 1.

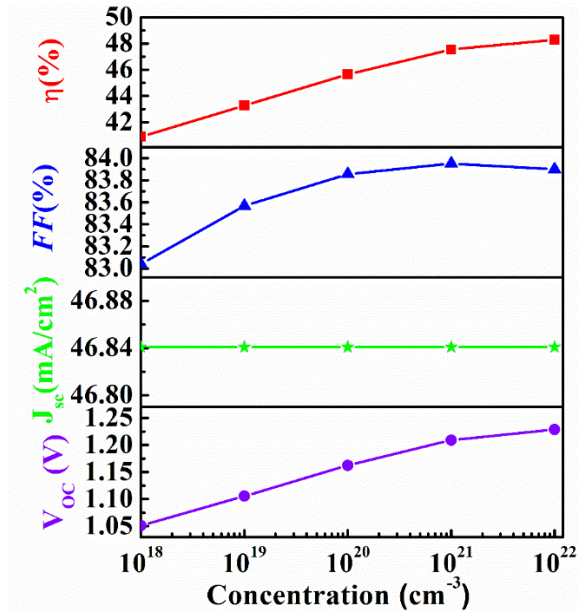

**Fig. 7.** Variation of the photovoltaic parameters with the doping density of BaSi$_2$.

With an increase in $N_A$, $J_{SC}$ remains constant, whereas all other parameters increase according to the increase in $V_{OC}$. The reverse saturation current reduces with an increase in $N_A$ of the BSF layer, resulting in an increase in the photo-generated voltage (Ali et al., 2019; Moon et al., 2020a). Khan and Suemasu (2017) have reported that the acceptor-type doping density in BaSi$_2$ can be done upto $10^{20}$ cm$^{-3}$ at room temperature (RT) if doped with boron (B). In this work, such a heavily-doped BaSi$_2$:B material is used as the BSF layer.

*3.6. Light-biased J-V characteristics and temperature coefficeints of the designed HPGe-based solar cells*

Taking thicknesses of 0.1 μm for *n*, *p$^+$* and 2 μm for *p* region, doping (cm$^{-3}$) of $10^{18}$ for *n*, $10^{17}$ for *p* and $10^{20}$ for *p$^+$* region and bulk defect (cm$^{-3}$) of $10^{14}$ for *n*, *p$^+$* and $10^6$ for *p* region, respectively, HPGe-based single homojunction (n-HPGe/p-HPGe), single heterojunction (n-CdS/p-HPGe) and double heterojunction (n-CdS/p-HPGe/p$^+$-BaSi$_2$) solar cells are simulated under a 1 sun illumination of AM 1.5G spectrum at 300 K. Fig. 8 demonstrates the *J-V* characteristic curves for these cases and the PV parameters of the coeesponding devices are shown in Table 2. The HPGe homojunction solar cells provide an efficiency of ~7.64% which is consistent with the practical Ge solar cells (Posthuma et al., 2007). On the other hand, HPGe heterojunction solar cell with CdS yields a PCE of ~21% which is also consistent with the Shockley-Queisser (SQ) detailed-balance limit for a solar cell whose absorber layer has a bandgap of 0.66 eV (Shockley et al., 1961). However, the efficiency of the HpGe-based solar cells increases to ~45.65% when BaSi$_2$ BSF layer is added to this heterojunction, which is almost consistent with the detailed-balance limit of ~42% for a double heterojunction solar cell (Vos, 1980).

Fig. 9 shows the histogram of the corresponding PV outputs parameters of HPGe-based single homojunction (n-HPGe/p-HPGe), single heterojunction (n-CdS/p-HPGe) and double heterojunction (n-CdS/p-HPGe/p$^+$-BaSi$_2$) solar cells. it is to be noted that the density of bulk defects in the Ge absorber layer significantly hampers the solar cells performance. For the demonstration, the double heterojunction is simulated freshly taking the absorber defect density of $10^{12}$ cm$^{-3}$, which implies the case of a highly defective low-purity Ge (LPGe). Also, its impact on the *J-V* curve is shown in Fig. 8 and the histograms of Fig. 9 displays the effect on PV parameters.

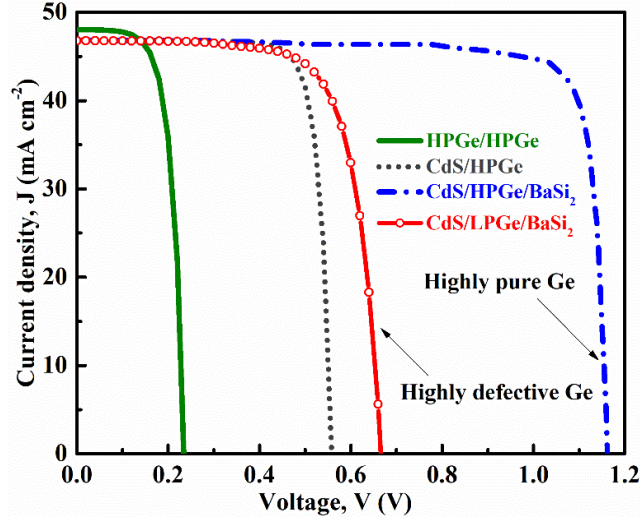

**Fig. 8.** Light *J-V* characteristics for HPGe-based homo- and hetero-junction solar cells.

**Table 2** The output PV parameters of HPGe-based solar cells obtained at 300 K un der light-bias.

| Case study | Junction type | Presence of BSF layer | Built-in voltage, $\psi$ (V) | $V_{OC}$ (V) | $J_{SC}$ (mA.cm$^{-2}$) | FF (%) | PCE (%) |
|---|---|---|---|---|---|---|---|
| n-HPGe/p-HPGe | Homo | NO | 0.49 | 0.24 | 48.03 | 67.60 | 7.64 |
| n-CdS/p-HPGe | Hetero | NO | 1.05 | 0.56 | 46.82 | 81.15 | 21.22 |
| n-CdS/p-HPGe/p$^+$-BaSi$_2$ | Hetero | YES | 1.70 | 1.16 | 46.84 | 83.87 | 45.65 |

It is evident from Fig. 8 and 9 (a) that HPGe in its single *n-p* homojunction is not feasible for high performance because of very low open circuit voltage and hence PCE, although a high photocurrent is generated. For the single *n-p* heterojunction, it is in consort to some extent with the solar cells currently available in the PV market and it becomes the most promising in the *n-p-p$^+$* double-heterojunction structure having a BSF. With the BaSi$_2$ BSF layer, the $V_{OC}$ of the heterojunction is improved significantly as percieved from Fig. 8 and 9 (b) due to an enhanced built-in voltage ($V_{bi}$). Thus, HPGe could be a potential photonic material when associated with CdS and BaSi$_2$ partners.

Fig. 9 (b) also makes it clear that although the double heterojunction solar cell possesses the same $V_{bi}$ for both the defect densities of either $10^6$ (highly pure state) or $10^{12}$ (highly defective state) cm$^{-3}$, the $V_{OC}$ is greatly hampered by the higher defect state and decreases from 1.16 to 0.667 V. As a result, all other parameters excluding the current tend to decrease tremendously as shown in Fig. 9 (a). It is found that about 1.74 times decrement in $V_{OC}$ reduces the PCE to almost half of the initial PCE.

It is realized that when Ge defect density is excessively high ($\geq 10^{15}$ cm$^{-3}$), all the solar cell structures are too incapable to give minimum performance. Overall, the grown-in defects in Ge must have to be controlled to achieve high-quality Ge for the best performance of the proposed double heterojunction solar cell.

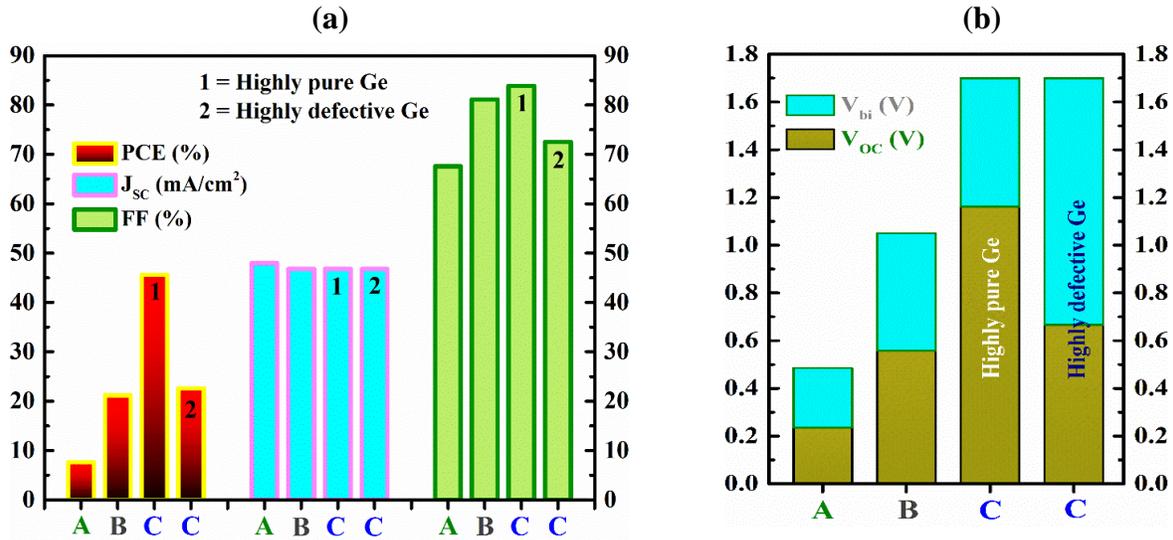

**Fig. 9.** PV parameters of HPGe-based solar cells (A = single homo-, B = single hetero-, C = double hetero-juncton as plotted in Fig. 8).

However, the degradation of the PV parameters of the proposed HJSC as a function of cell operating temperature is also analyzed. In terms of PCE, temperature coefficient, $T_C$ at standard test conditions (STC) is calculated using Eq. (6) of the article (Moon et al., 2020b). The proposed solar cell offers a $T_C$ of -0.1206% K$^{-1}$ asserting a suitable thermal stability of the device.

## 4. Conclusion

High-purity germanium (HPGe)-based novel $n$-CdS/$p$-HPGe/$p^+$-BaSi$_2$ heterojunction solar cells (HJSCs) have been designed and simulated using SCAPS-1D software. The structure of the device has been optimized for the highest efficiency of the solar cells. It is revealed that the defect density of HPGe has a siginificant effect on the open-circuit voltage compared to the other parameters of HPGe as well as partner layers. The PCE of the devices reduces from 45.65% to 31.54% when the defect density of HPGe increases from $10^6$ to $10^{10}$ cm$^{-3}$, respectively. The built-in potential of the device is calculated to ~1.7 V. The highest PCE of the solar cells has been achieved to ~45.65% for a total device thickness of 2.2 µm with a high $V_{OC}$ of 1.16 V despite the low band gap of the HPGe absorber layer. The optimized FF and $J_{SC}$ are 83.87% and 46.84 mA cm$^{-2}$, respectively. The current study explores the possibility of HPGe as a promising photonic material via the newly designed HPGe-based $npp^+$ double-HJSC for large-scale photovoltaic applications in space and multi-kilowatt systems of the PV industries. Overall, this article reveals the design guidelines towards high-efficiency solar cells to be fabricated in the near future.

## Acknowledgments

The authors highly appreciate Professor Marc Burgelman of the department of Electronics and Information Systems, University of Gent, Belgium for providing SCAPS software package.

*Corresponding author:
E-mail: jak_apee@ru.ac.bd

## Funding

This research did not receive any specific grant from funding agencies in the public, commercial, or not-for-profit sectors.

## Declaration of interests

The authors declare that they have no known competing financial interests or personal relationships that could have appeared to influence the work reported in this paper.

Gupta, S., Magyari-Köpe, B., Nishi, Y., Saraswat, K.C., 2013. Achieving direct band gap in germanium through integration of Sn alloying and external strain. J. Appl. Phys. 113, 073707. https://doi.org/10.1063/1.4792649.

Haller, E.E., Hansen, W.L., Goulding, F.S., 1972. Nuclear science symposium; Miami Beach, Florida, USA CONF-721202-12 481.

Haller, E.E., Pearl, P.L., Hubbard, G.S., Hansen, W.L., 1979. Deep Level Transient Spectroscopy of High-Purity Germanium Diodes/Detectors, IEEE Trans. Nucl. Sci. 26, 265 - 270. https://doi.org/10.1109/TNS.1979.4329643.

Hansen, W.L., Haller, E.E., 1981. Fabrication Techniques for Reverse Electrode Coaxial Germanium Nuclear Radiation Detectors, IEEE Trans. Nucl. Sci. 28, 541-543.

Hara, K.O., Nakagawa, Y, Suemasu, T., Usami, N., 2016. Simple vacuum evaporation route to $BaSi_2$ thin films for solar cell applications. Procedia Engineering 141. https://doi.org/10.1016/j.proeng.2015.08.1103.

Hou, Y., Aydin, E., Bastiani, M.D., Xiao, C., Isikgor, F.H., Xue, D.J., Chen, B., Chen, H., Bahrami, B., Chowdhury, A.H., Johnston, A., Baek, S.W., Huang, Z., Wei, M., Dong, Y., Troughton, J., Jalmood, R., Mirabelli, A.J., Allen, T.G., Kerschaver, E.V., Saidaminov, M.I., Baran, D., Qiao, Q., Zhu, K., Wolf, S.D., Sargent, E.H., 2020. Efficient tandem solar cells with solution-processed perovskite on textured crystalline silicon, Science 367, 1135-1140. https://doi.org/10.1126/science.aaz3691.

Kasahara, K., Hossain, J., Harada, D., Ichikawa, K., Ishikawa, R., Shirai, H., 2018. Crystalline-Si heterojunction with organic thin-layer (HOT) solar cell module using poly(3,4-ethylenedioxythiophene):poly(styrene sulfonate)(PEDOT:PSS). Sol. Energy Mater. Solar Cells 181, 60-70. https://doi.org/10.1016/j.solmat.2017.10.016.

Kayesh, M.E., Matsuishi, K., Kaneko, R., Kazaoui, S., Lee, J.J., Noda, T., Islam, A., 2019. Coadditive Engineering with 5-Ammonium Valeric Acid Iodide for Efficient and Stable Sn Perovskite Solar Cells. ACS Energy Lett. 4, 278-284. https://doi.org/10.1021/acsenergylett.8b02216.

Khan, M.A., Suemasu, T., 2017. Donor and acceptor levels in impurity‐doped semiconducting $BaSi_2$ thin films for solar‐cell application. Phys. Status Solidi A 214, 1700019. https://doi.org/10.1002/pssa.201700019.

Kose, S., Atay, F., Bilgin, V., Akyuz, I., Ketenci, E., 2010. Optical characterization and determination of carrier density of ultrasonically sprayed CdS:Cu films. Appl. Surf. Sci. 256, 4299-4303. https://doi.org/10.1016/j.apsusc.2010.02.018.

Landolt-Börnstein, 2002. Group III Condensed Matter: Germanium (Ge), intrinsic carrier concentration, in: Madelung, O., Rössler, U., Schulz, M. (Eds.), Group IV Elements, IV-IV and III-V Compounds. Part b - Electronic, Transport, Optical and Other Properties. Springer-Verlag Berlin Heidelberg. pp. 1–3. https://doi.org/10.1007/10832182_503.

Li, Y., Guo, H., Yao, Y., Dutta, P., Rathi, M., Zheng, N., Gao, Y., Sun, S., Ryou, J.H., Ahrenkiel, P., Selvamanickam, V., 2018. Defect reduction by liquid phase epitaxy of germanium on single-21